\documentclass[twocolumn,showpacs,preprintnumbers,amsmath,amssymb,superscriptaddress,floatfix]{revtex4}

\usepackage{graphicx}
\usepackage{dcolumn}
\usepackage{bm}
\usepackage{latexsym}

\begin{document}


\title{The Dual Meissner Effect
 and
%
%
 Magnetic Displacement Currents
}

\author{Tsuneo Suzuki}
\affiliation{Institute for Theoretical Physics, Kanazawa University, 
Kanazawa 920-1192, Japan}
\affiliation{RIKEN, Radiation Laboratory, Wako 351-0158, Japan
}%
\author{Katsuya Ishiguro}
\affiliation{Institute for Theoretical Physics, Kanazawa University,
Kanazawa 920-1192, Japan}
\author{Yoshihiro Mori} 
\affiliation{Institute for Theoretical Physics, Kanazawa University,
Kanazawa 920-1192, Japan}
\author{Toru Sekido}
\affiliation{Institute for Theoretical Physics, Kanazawa University,
Kanazawa 920-1192, Japan}


\begin{abstract} 
The dual Meissner effect is observed without monopoles in quenched
 $SU(2)$ QCD with Landau gauge-fixing. Magnetic displacement currents 
that 
are time-dependent Abelian magnetic fields 
act as
solenoidal currents squeezing Abelian electric fields.  Monopoles are
 not always necessary 
for
the dual Meissner effect. 
%
%
%
A mean-field calculation suggests that the dual Meissner effect through 
the  mass generation 
of the Abelian electric field is related to a gluon condensate  $\langle A^a_{\mu}A^a_{\mu}\rangle \neq 0$ of mass dimension 2.  
\end{abstract}

\maketitle

The understanding of the
color confinement mechanism is 
an important problem that is yet to be solved.
It is believed that the dual Meissner effect is the mechanism\cite{tHooft:1975pu, Mandelstam:1974pi}.
However, the cause of the dual Meissner effect has not yet been clarified. 
A possible 
cause 
is 
the appearance of 
magnetic monopoles 
after projecting $SU(3)$ QCD to an Abelian $U(1)^2$ theory by 
partial gauge fixing\cite{tHooft:1981ht}. If such monopoles condense, 
the dual Meissner effect could explain the color confinement.
In fact,  
an Abelian projection adopting a special gauge called 
maximally 
Abelian gauge (MA)\cite{suzuki-83,kronfeld} leads us to interesting
results\cite{AbelianDominance,Reviews} 
that
support 
importance 
 of 
%
%
 monopoles.

Now, the following question arises: What happens when other general
Abelian projections are adopted or when no Abelian projection is adopted?
For example, consider an Abelian projection diagonalizing Polyakov loops. Monopoles exist in the continuum limit at a point where  eigenvalues of Polyakov loops are degenerate\cite{tHooft:1981ht}.
However, it can be easily shown
that such a point 
moves
only in 
a 
time-like direction. 
In other words,
there exist
only time-like monopoles 
that 
do not contribute to the string tension
%
%
\cite{maxim}. 
Let us discuss 
%
another simple case of 
the 
Landau gauge. 
%
%
Vacuum configurations are 
%
%
smooth and 
%
%
monopoles 
arising
from singularities 
do not 
exist. 
%
In these cases,
monopole condensation 
%
does
 not occur. We have to find 
%
%
a more general
 confinement mechanism
%
that
is realized in QCD. 
%

This study shows
that the dual Meissner effect in an Abelian sense works 
well 
even when monopoles do not exist.
Monte-Carlo simulations of quenched $SU(2)$ QCD 
in
%
the
Landau gauge 
are adopted. 
Instead of monopoles, time-dependent Abelian magnetic fields regarded as
magnetic displacement currents 
squeeze the
Abelian electric fields. The dual Meissner effect 
implies
the dual London equation and 
mass generation of the Abelian electric fields 
that 
may be related to
%
the
existence of a dimension 
2 
gluon condensate. 
The
present numerical results are not perfect
 since the continuum limit, 
infinite-volume 
limit,
 and 
%
gauge-independence 
have not yet been examined.
These
discussions use 
the 
Abelian components 
based 
only on the 
assumption that Abelian components are dominant in 
infrared QCD (Abelian
dominance\cite{AbelianDominance,Reviews,greensite-96,Cea:1995zt})
; however, this assumption has not yet been clarified. 
Nevertheless, 
the
authors 
believe that
the 
 results 
obtained here
 are very 
interesting,  
since they show for the first time 
that 
the Abelian dual Meissner effect 
works
in lattice non-Abelian QCD without 
%
%
requiring
%
monopoles 
from a singular gauge transformation\cite{faber}. The gauge adopted here is 
the
%
simple
one 
and is used 
only 
for the purpose of 
obtaining
 smooth configurations. 
%
%
Hence,
these 
%
results 
suggest 
that
the Abelian dual Meissner effect is the 
actual
universal mechanism of color confinement. 
%
Moreover, 
the 
probable 
relation 
between
the Abelian dual Meissner effect 
and
the dimension 
2 
gluon condensate sheds 
%
%
new light on the importance of the gluon condensate
\cite{zakharov,kondo,dudal,arriola,boucaud,slavnov,li}. 
%
%

\begin{figure}[b]
\includegraphics[height=6cm, width=8.5cm]{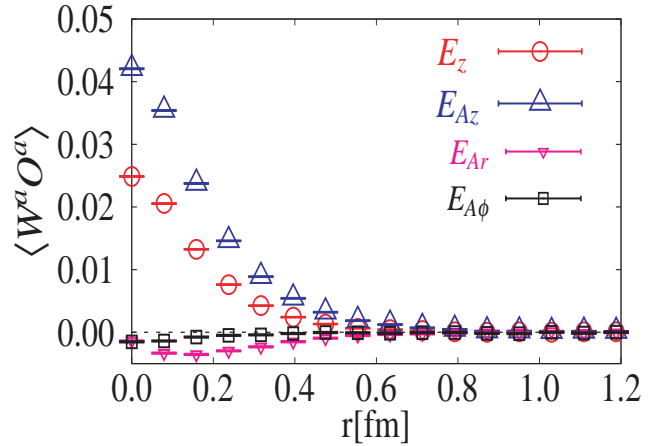}
\caption{\label{fig_1} Abelian $\vec{E}_A$ and non-Abelian $\vec{E}$ electric field  profiles in Landau gauge. $W(R\times T= 6\times 6)$ is used.}
\end{figure}

\begin{figure*}
\includegraphics[height=5.5cm]{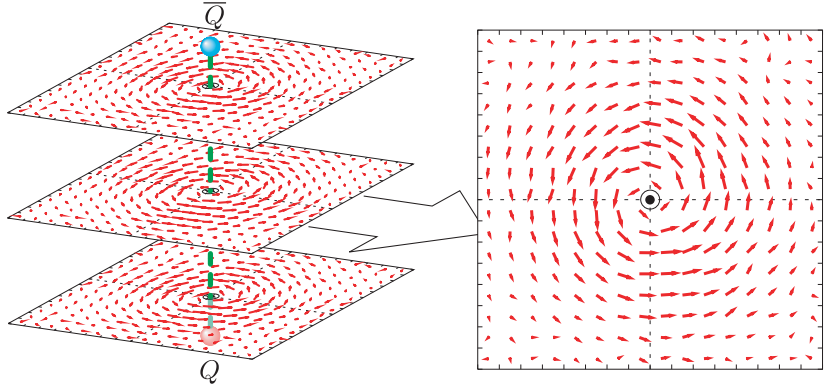}
\caption{\label{fig_2}Magnetic displacement currents in Landau gauge as a solenoidal current.}
\end{figure*}

%
We 
use 
the following 
improved gluonic action found by Iwasaki\cite{iwasaki}
as a lattice $SU(2)$ QCD:
\begin{eqnarray}
S = \beta \left\{c_0 \sum Tr (plaquette) + c_1 \sum Tr (rectangular) \right\},\nonumber
\end{eqnarray}
which enables us to obtain
better scaling behaviors of physical quantities. 
%
The mixing parameters are fixed as 
$c_0 + 8c_1 = 1$ and  $c_1=-0.331$.

In order to directly
measure 
the 
correlations of gauge-variant electric and magnetic 
fields, 
we adopt 
the 
simplest 
gauge, 
called  Landau 
gauge, 
which maximizes
$\sum_{s,\mu} Tr [ U_{\mu}(s) + U_{\mu}^{\dagger}(s)]$.
After the gauge fixing, we 
attempt
to measure electric and magnetic flux distributions by evaluating correlations of Wilson loops and field strengths. For comparison, we also 
discuss  
%
the 
MA gauge where 
$\sum_{s,\mu} Tr [ U_{\mu}(s)\sigma_3U_{\mu}^{\dagger}(s)\sigma_3]$
is maximized.
In order to obtain
a good signal to noise ratio,  the APE smearing technique\cite{APE}
is used when evaluating 
the 
Wilson loops $W(R, T)=W^0+iW^a\sigma^a$.

The measurements
of the string tension 
%
%
are used to
fix the scale when we use $\sqrt{\sigma}=440$[MeV].
We adopt a coupling constant $\beta=1.2$ in which the lattice distance
$a(\beta=1.2)$ is $0.0792(2)$[fm]. 
This 
%
%
lattice distance
is chosen 
based on the comparison between our results and those of the
MA gauge\cite{bali-96,Koma:2003gq}.
The lattice size is $32^4$ and after 5000 
thermalizations, 
we have 
%
%
taken
5000 thermalized configurations per 
100 sweeps for measurements.

Non-Abelian electric and magnetic fields  are defined from $1\times 1$
plaquette 
$U_{\mu\nu}(s)=U^0_{\mu\nu}+iU^a_{\mu\nu}\sigma^a$,
similar to
Ref.\cite{bali-94}: 
\begin{eqnarray}
E^a_k(s)&\equiv& \frac{1}{2}(U^a_{4k}(s-\hat{k})+U^a_{4k}(s)), \nonumber\\
B^a_k(s)&\equiv& \frac{1}{8}\epsilon_{klm}(U_{lm}^a(s-\hat{l}-\hat{m})\nonumber \\
&&+U_{lm}^a(s-\hat{l})+U_{lm}^a(s-\hat{m})+U_{lm}^a(s)). \nonumber
\end{eqnarray}
We also define 
the 
Abelian electric ($E_{Ai}^a$) and magnetic fields ($B_{Ai}^a$) 
in a similar manner
using 
the 
Abelian plaquettes $\theta_{\mu\nu}^a(s)$ defined 
with the
link variables $\theta_{\mu}^a(s)$: 
\begin{eqnarray}
\theta_{\mu\nu}^a(s)\equiv\theta_{\mu}^a(s)+\theta^a_{\nu}(s+\hat{\mu})-\theta_{\mu}^a(
s+\hat{\nu})-\theta^a_{\nu}(s),
\end{eqnarray}
where $\theta_{\mu}^a(s)$ is given by 
$U_{\mu}(s)=\exp(i\theta_{\mu}^a(s)\sigma^a)$.
In 
the 
MA gauge, the Abelian link variable $\theta_{\mu}^{MA}(s)$ is defined by a phase of the diagonal part of a non-Abelian link field:
\begin{eqnarray*}
U_{\mu}^0(s)&=&\sqrt{1-|c_{\mu}(s)|^2}\cos\theta^{MA}_{\mu}(s),\\
U_{\mu}^3(s)&=&\sqrt{1-|c_{\mu}(s)|^2}\sin\theta^{MA}_{\mu}(s).
\end{eqnarray*}
Since the off-diagonal part $|c_{\mu}(s)|$ is small\cite{AbelianDominance},  $\theta_{\mu}^{MA}(s)\sim \theta^3_{\mu}(s)$ in MA gauge. 
As a source corresponding to a static quark and antiquark pair,
we 
adopt
%
only non-Abelian Wilson loops 
in this study.

First, 
we show 
Abelian and non-Abelian electric  
field
 profiles around a 
quark
pair
in 
the 
Landau gauge
in Fig.\ref{fig_1}. 
The profiles are 
%
studied 
mainly
on 
a
perpendicular plane at the midpoint between the 
two
quarks.
It must be noted
that 
the electric fields perpendicular to the $Q\bar{Q}$ axis are found to be
negligible. It is very interesting to 
observe
from Fig.\ref{fig_1} that 
the
Abelian electric field $E_{Az}$ 
%
defined here is 
also squeezed
although short-range Coulombic contributions are different\cite{gluonmass}.
%
%
%
It is, thus, essential to ascertain the reason for the squeezing of 
the Abelian flux.

Let us 
now 
discuss 
only the
flux distributions of Abelian  
fields.
It is 
numerically verified\cite{sardinia} 
that  
no DeGrand-Toussaint monopoles\cite{DeGrand:1980eq} 
are present.
Hence,
the Abelian fields satisfy 
the simple Abelian Bianchi identity kinematically, as demonstrated below:
\begin{eqnarray}
\vec{\nabla}\times\vec{E}_{A}^a=\partial_{4}\vec{B}_{A}^a, \hspace{1cm}
\vec{\nabla}\cdot\vec{B}_{A}^a=0.\label{BI1}
\end{eqnarray}
In the case of MA gauge, there 
exist
additional monopole current $(\vec{k}, k_4)$ contributions:
\begin{eqnarray}
\vec{\nabla}\times\vec{E}^{MA}=\partial_{4}\vec{B}^{MA}+\vec{k}, \hspace{1cm}
\vec{\nabla}\cdot\vec{B}^{MA}=k_4.\label{BI2}
\end{eqnarray}
Here,
$\vec{E}^{MA}$ and $\vec{B}^{MA}$ are defined in terms of plaquette variables 
$\theta_{\mu\nu}^{MA}(s)\ \  (\textrm{mod}\ \ 2\pi)$ 
that
are constructed 
with
 $\theta^{MA}_{\mu}(s)$. 
 
The Coulombic electric field 
arising
from the static source is 
expressed in terms of the gradient of a scalar potential in the 
lowest perturbation theory.
Hence,
it 
contributes neither
to the curl of the Abelian electric field nor to the Abelian magnetic field in 
the 
Abelian Bianchi 
identity,
Eq.(\ref{BI1}).
According to the
dual Meissner 
effect,
the squeezing of the electric flux occurs due to 
the 
cancellation of 
Coulombic electric fields and those 
due to the
solenoidal magnetic currents. In the case of  
the
MA gauge, magnetic monopole currents $\vec{k}$ 
serve as 
the solenoidal current\cite{Singh:1993jj,bali-96,Koma:2003gq}. 

\begin{figure}
\includegraphics[height=6cm, width=8.5cm]{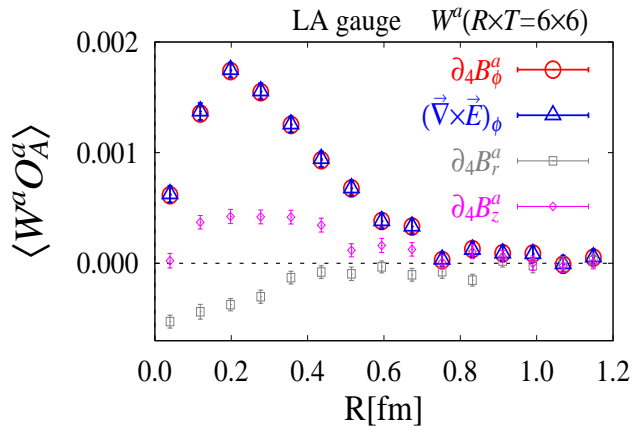}
\caption{\label{fig_3}Curl of Abelian electric fields and magnetic displacement currents around a static quark pair in Landau gauge.}
\end{figure}

\begin{figure}
\includegraphics[height=6cm,  width=8.5cm]{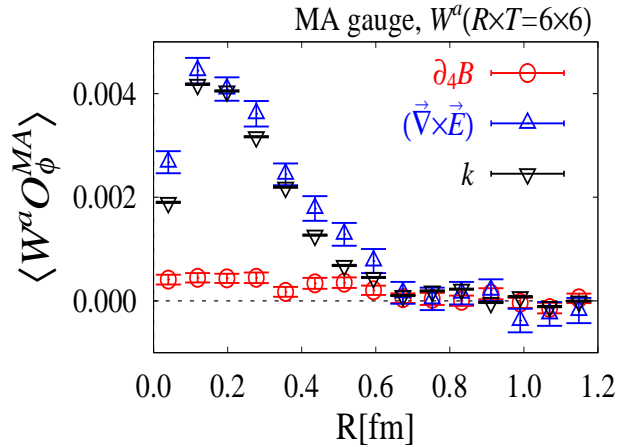}
\caption{\label{fig_4}Curl of Abelian electric fields, 
monopole currents,
and magnetic displacement currents around a static quark pair in MA gauge.}
\end{figure}

What
happens in a smooth gauge like 
the 
Landau gauge where monopoles do not exist?
It can be seen from
Eq.(\ref{BI1})
that
only 
$\partial_4\vec{B}_{A}$, which is
regarded as a magnetic displacement  
current, can
play the role of the solenoidal  current. It is very interesting to see
Fig.\ref{fig_2}  
that demonstrates the occurrence of this phenomenon
in Landau gauge. Note that the solenoidal current 
is in 
a direction 
that squeezes
the Coulombic electric field. Let us 
also see
the 
%
$R$ dependence
 shown in Fig.\ref{fig_3}. The 
%
 components of the magnetic displacement current $\partial_4B_{Ar}$ and
 $\partial_4B_{Az}$ 
do not vanish;
however,
they are 
extremely suppressed.  
In comparison, we 
present
the
 case of 
the 
MA gauge in Fig.\ref{fig_4}. 
Here,
$\partial_4B_{A\phi}$ is found to be 
numerically negligible
as already expected  from the 
cited literatures\cite{Singh:1993jj,Cea:1995zt,bali-96}. 
Instead monopole currents are 
found to circulate\cite{Singh:1993jj,bali-96,Koma:2003gq}. 
In this case, $k_r$ is non-vanishing, although it is also suppressed
 in comparison with $k_{\phi}$. $k_z$ is almost zero.
 The authors 
believe
that 
the 
non-vanishing of the radial and
%
%
  $z$ components of $\partial_4\vec{B}$ in 
the 
Landau gauge and 
$k_r$ 
in 
the 
MA gauge is  due to lattice artifacts and 
the small size of
the Wilson loop 
used
here.
It is interesting that the shapes of $\partial_4B_{A\phi}$ in 
the 
Landau gauge and $k_{\phi}$ in 
the 
MA gauge 
appear to be similar even though
the strengths are different. 
These shapes
have a peak at almost the same 
distance 
at approximately
$0.2$[fm] and almost vanish 
at approximately
$0.7$[fm].  

\begin{figure}
\includegraphics[height=6cm, width=8.5cm]{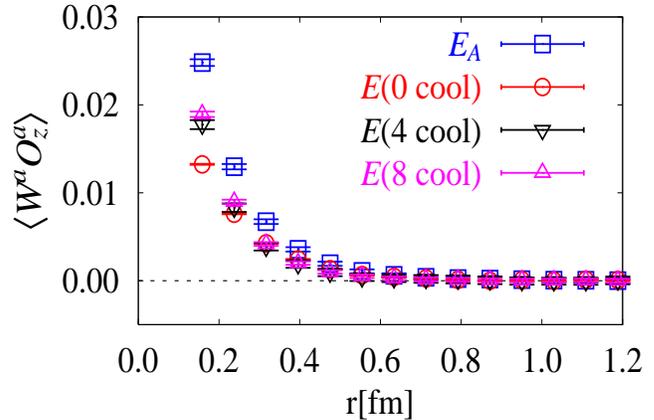}
\caption{\label{fig_5}Profiles of non-Abelian electric field $E_z^a$
 after cooling. $E_{Az}^a$ 
represents
the Abelian electric field. $W(R\times T= 6\times 6)$ is used.}
\end{figure}

It can be speculated whether the above consideration of Abelian fields
is sufficient to understand the non-perturbative confinement problem in
infrared QCD in terms of Abelian quantities\cite{AbelianDominance,Reviews,kondo-04}. 
Let us 
%
attempt to verify the
Abelian dominance in 
the 
Landau gauge using a controlled cooling \cite{cooling} under which  the string tension remains non-vanishing. 
It is found that
 the profile of the non-Abelian electric field $E_{z}^a$ tends to that
 of the Abelian one of $E_{Az}^a$ in the long-range 
region, 
as shown in Fig.\ref{fig_5}. This is consistent
with the 
previous
result\cite{Cea:1995zt} 
using
a different approach.

It 
has been
shown that the magnetic  displacement currents are important in the dual
Meissner effect when there are no monopoles. 
In such a case, 
how can 
the origin of the dual Meissner effect without 
monopole condensation
be understood?
The Abelian dual Meissner effect indicates the massiveness of the Abelian electric field as an asymptotic field:
\begin{eqnarray}
(\partial_{\rho}^2-m^2)\vec{E}_A\sim 0.\label{eqm-abel}
\end{eqnarray}
This leads us to  a dual London equation 
that 
is a key  to the dual Meissner effect. Let us evaluate the curl of the
magnetic  displacement current. Using Eq.(\ref{BI1}), we 
obtain
\begin{eqnarray}
\vec{\nabla}\times\partial_4\vec{B}_A&=&\vec{\nabla}(\vec{\nabla}\cdot\vec{E}_A)-\vec{\nabla}^2\vec{E}_A.\nonumber 
\end{eqnarray}
From Eq.(\ref{eqm-abel}), we 
obtain
the dual London equation:  
\begin{eqnarray}
\vec{\nabla}\times\partial_4\vec{B}_A\sim (\partial_4^2-m^2)\vec{E}_A.\label{London2}
\end{eqnarray}

Let us 
use the
simple mean-field approach developed by Fukuda\cite{fukuda-82}.
By neglecting the
gauge-fixing and Fadeev-Popov terms, we 
obtain the 
equation
of motion
$D_{\mu}^{ab}F_{\mu\nu}^b= 0$ 
and the (non-Abelian) Bianchi identity
$D_{\mu}^{ab}{}^*F_{\mu\nu}^b=0$.
By applying
$D$ operator to the Bianchi identity and using the Jacobi identity and
the equations of motion,  we 
obtain 
$(D_{\rho}^2)^{ab}F_{\mu\nu}^b=2g\epsilon^{abc}F_{\mu\alpha}^bF_{\nu\alpha}^c$.
Note that 
$(D_{\rho}^2)^{ab} = \partial_{\rho}^2\delta^{ab}+g\epsilon^{acb}(\partial_{\rho}A_{\rho}^c)+g^2(A_{\rho}^aA_{\rho}^b-\delta^{ab}(A_{\rho}^c)^2)$. 
Hence,
if $\langle A^a_{\mu}A^b_{\nu}\rangle =\delta^{ab}\delta_{\mu\nu}v^2\neq 0$, we see 
asymptotically that the electric fields become massive $(\partial_{\rho}^2-m^2)E^a_k\sim 0$
with $m^2=8g^2v^2$.
%
%
Now, 
the Abelian electric field is also massive asymptotically
$(\partial_{\rho}^2-m^2)E^a_{Ak}\sim 0$.
Hence, 
the dual London equation (\ref{London2}) is  obtained.

The importance of the dimension 
2 
gluon condensate has been stressed by Zakharov and his collaborators\cite{zakharov} and also 
by many authors 
in Refs.\cite{kondo,dudal}. Recent discussions on the value of the gluon
condensate 
can be
seen in Ref.\cite{arriola}. Some 
discuss the mass generation of the gluon propagator
\cite{gracey}. However, it is found\cite{li} that the gluon propagator
has no on-mass-shell poles in the presence of the dimension 
2  
gluon condensate. This does not 
directly imply 
that the electric field propagator  has  no on-mass-shell 
pole
since a gluon field $\vec{A}$ and an electric field $\vec{E}$ are different canonical variables\cite{fukuda-82}. 
The lattice study of gluon propagators  in 
the 
Landau gauge 
demonstrates the 
importance of the dimension 
2  
condensate and its relevance to instantons\cite{boucaud}.     
Although the operator of the gluon condensate is 
gauge variant,
there have been various discussions concerning  gauge-invariant contents in the 
expectation value of the operator\cite{zakharov,kondo,dudal,slavnov}.

%



The numerical simulations of this work were 
performed
using RSCC computer clusters in 
RIKEN. The authors would like to thank RIKEN for their 
computer facilities. 
T.S. is supported by JSPS Grant-in-Aid for Scientific Research on Priority Areas 13135210 and (B) 15340073.

\begin{thebibliography}{99}
\bibitem{gluonmass}
The gluon mass determined from  Fig.\ref{fig_1} is roughly 
$1.35$
[GeV]. 
However,
the distance $R$ between the quark pair in Fig.\ref{fig_1} is only 
$R= 0.48$
[fm] 
and it is too short to 
correctly fix
the penetration length and the gluon mass in the confinement force. 
In fact,
the penetration length observed 
increases
up to $R= 0.8$
[fm]
which is
consistent with the results in Ref.\cite{bali-94}. The details will be published elsewhere.
\end{thebibliography}

\end{document}